\documentclass[10pt]{article}

\usepackage{amsmath}

\begin{document}


\begin{center}
{\Large {\bf Manifold of density matrices}}
\end{center}

\hfil\break

\begin{center}
{\bf Aik-meng Kuah\footnote{kuah@physics.utexas.edu}, E.C.G. Sudarshan \\}
{\it Department of Physics \\
University of Texas at Austin \\
Austin, Texas 78712-1081}\\
\end{center}

\begin{center}
July 30 2003
\end{center}

\begin{abstract}

We show that the manifold of density matrices can be derived from $CP^{N^2-1}$ by the action of $SU(N)$.  We give some preliminary observations on the structure of this manifold. 

\end{abstract}

\hfil\break
\hfil\break

\pagebreak


\section{Introduction}

The manifold of pure quantum states is well known and well described, but the generalized quantum states, the set of density matrices (hermitian positive matrices of unit trace) remain not very well understood.

In this paper we set out to find the manifold of density matrices.  We show that the manifold of density matrices can be derived from the $CP^{N^2-1}$ by the action of $SU(N)$.

\section{Manifold of density matrices}

The technique of purification  ~\cite{purification} allows us to describe $N \times N$ density matrices in terms of $N^2$ dimensional pure states.  Any $N \times N$ density matrix has a canonical decomposition:

\begin{equation}
\rho = \sum_i^N \rho_i |i^A\rangle\langle i^A|
\end{equation}

where $\rho_i$ and $|i^A\rangle$ are the eigenvalues and eigenvectors, respectively.  

Using the purification procedure, we can always find a pure state in a $N^2$ dimensional Hilbert space, given by its Schmidt decomposition:

\begin{equation}
|\psi\rangle = \sum_i^N \sqrt{\rho_i}|i^A\rangle|i^B\rangle
\end{equation}

where the partial trace over the ancillary system yields the original density matrix:

\begin{equation}
Tr_B[|\psi\rangle\langle\psi|] = \sum_i^N \rho_i |i^A\rangle\langle i^A| = \rho
\end{equation}

Note that we have labelled the states in the original space of the density matrix with the superscript $A$ and the ancillary system with the superscript $B$.  Also note that $|i^B\rangle$ forms an orthornormal basis.

We can consider this as a many-to-one mapping from the set of pure states in $N^2$ dimensions to density matrices in $N$ dimensions.  The following lemma allows us to determine just how many pure states would correspond to the same density matrix. 

Lemma 1:  Given 2 pure states in a $N^2$ dimensional Hilbert space $|\phi\rangle$ and $|\psi\rangle$, with the same partial trace:

\begin{equation}
Tr_B[|\phi\rangle\langle\phi|]= Tr_B[|\psi\rangle\langle\psi|]
\end{equation}

then there exists a unitary transformation $v \in SU(N)$ such that $I^A \otimes v^B |\psi\rangle = |\phi\rangle$.

To prove Lemma 1, let us write the 2 states in their Schmidt decomposition:

\begin{equation}
|\psi\rangle = \sum_k^\mu q_k|q_k^A\rangle|q_k^B\rangle
\end{equation}

\begin{equation}
|\phi\rangle = \sum_k^\nu p_k|p_k^A\rangle|p_k^B\rangle
\end{equation}

Their partial traces are equal:

\begin{equation}
\sum_k^\mu q_k^2 |q_k^A\rangle\langle q_k^A| = \sum_k^\nu p_k^2 |p_k^A\rangle\langle p_k^A|
\end{equation}

iff $\mu=\nu$, $q_k^2 = p_k^2$ and if no values $q_k^2$ are degenerate then $|q_k^A\rangle = |p_k^A\rangle$ upto a phase.

In the non-degenerate case, the only freedom is in specifying the basis $\{|q_k^B\rangle\}$ and $\{|p_k^B\rangle\}$, which is given by a unitary transformation in the $B$ system.

If a degeneracy exists between $q_z^2$ for particular values of $z$, then the states $|q_z^A\rangle$ and $|p_z^A\rangle$ can differ by an unitary transform.  We will show that this unitary transformation can be equivalently performed in the $B$ subspace.  Consider the subspace of $|\psi\rangle$ for which the Schmidt coefficients are degenerate:

\begin{equation}
\sum_z C |q_z^A\rangle|q_z^B\rangle
\end{equation}

A unitary transformation $R$ within the space spanned by $|q_z^A\rangle$ gives:

\begin{equation}
\sum_z C R|q_z^A\rangle|q_z^B\rangle = \\
\sum_z C \sum_{z'} R_{zz'}|q_{z'}^A\rangle|q_z^B\rangle = \\
\sum_z C \sum_{z'} |q_z^A\rangle R_{z'z}|q_{z'}^B\rangle = \\
\sum_z C |q_z^A\rangle R^T |q_z^B\rangle
\end{equation}

Therefore we see that $R \otimes I |\psi\rangle = I \otimes R^T |\psi\rangle$.

And finally to complete the proof, any complex phase on the Schmidt coefficients can also be introduced by a unitary transformation on the $B$ subspace. 

With lemma 1, we see that $SU(N)$ orbits within the manifold of $N^2$ dimensional pure states $CP^{N^2-1}$ correspond to unique $N \times N$ density matrices.  This is a bijection between the set of $N \times N$ density matrices and these orbits on the manifold $CP^{N^2-1}/SU(N)$.

\section{Geometry of the manifold of density matrices}

Let us consider the action of $SU(N)$ on the manifold $CP^{N^2-1}$ using the following:

\begin{equation}
I^A \otimes U^B \sum_k^\mu q_k|q_k^A\rangle|q_k^B\rangle
\end{equation}

where $U^B \in SU(N)$ acts on a state in $CP^{N^2-1}$ written in its Schmidt decomposition.

This action is not free, therefore we need to classify the stabilizers of the states.  For a state with Schmidt number $\mu$, the orthornormal states $|q_k^B\rangle$ would span a $\mu$ dimensional subspace, so the stabilizer of this state is the subgroup $U(N-\mu)$.  This gives rise to different dimensional strata, identified by the Schmidt number (or equivalently the density matrix rank) $\mu$.  

The dimension of $CP^{N^2-1}$ is $2N^2-2$ and the dimension of $SU(N)$ is $N^2-1$.  This gives us the expected dimension of $N^2-1$ for the density matrices.  For states with maximal Schmidt number $\mu=N$, the stabilizer is the trivial group $1$, the dimension of the stratum is $N^2-1$.  As the Schmidt number decreases, the stabilizer group increases, and the dimension of the stratum decreases.  For rank $\mu$ matrices, the stratum is $\mu(2N-\mu)-1$ dimensional.  The lowest dimensional stratum is $2N-2$, which is the set of pure states, and which is also the manifold $CP^{N-1}$.

Each stratum is also a convex covering for the higher dimensional stratum, as it is clear that a rank $\mu$ density matrix can always be decomposed in terms of rank $\mu-1$ density matrices.  The lowest stratum, of the set of pure states, is the overall convex covering for all density matrices.

We now see the set of density matrices as a layering/strata of surfaces of dimensions given by $\mu(2N-\mu)-1$, with $1 \geq \mu \geq N$, with the lowest dimensional stratum, a $2N-2$ dimensional surface forming the set of pure states.

\section{Geometry of N=2 density matrices}

As a simple example, let us consider the case where $N=2$, the manifold of density matrices is given by $CP^3/SU(2)$.  There are only 2 possible density matrix ranks, giving rise to only 2 strata.  The stratum of pure states forms the convex hull and is 2 dimensional.  The stratum of rank 2 matrices, the mixed states, is 3 dimensional.

The manifold is symmetric under $SU(2)$, which coincides with $SO(3) \times Z_2$.  Given the symmetry and structure, the only possible shape is a sphere.  This agrees with the well known Bloch sphere representation for qubit density matrices.

\section{Conclusions and comments}

In this paper, we have obtained an interesting relationship between the manifold of density matrices and the well known manifolds of pure states $CP^N$.  This allows us to discuss a mathematical manifold for density matrices which has not been available to us before.  With this, we hope to develop a greater understanding of quantum density matrices and their operators (dynamical maps).

\section{Acknowledgments}

We would especially like to thank Dr. Tamas Hausel for his help on actions on manifolds.  We would also like to thank Anil Shaji and Dr. Todd Tilma for fruitful discussions on this subject.


\end{document}